# Single photon emission from individual nanophotonic-integrated colloidal quantum dots

*Alexander Eich[1,2,3], Tobias C. Spiekermann[1,2,3], Helge Gehring[1,2,3], Lisa Sommer[1,2,3], Julian R. Bankwitz[1,2,3], Philip P.J. Schrinner[1,2,3], Johann A. Preuß[1,2], Steffen Michaelis de Vasconcellos[1,2], Rudolf Bratschitsch[1,2], Wolfram H. P. Pernice[1,2,3], Carsten Schuck[1,2,3, *]*

1 Institute of Physics, University of Münster, Wilhelm-Klemm-Str. 10, 48149 Münster, Germany
2 Center for Nanotechnology (CeNTech), Heisenbergstr. 11, 48149 Münster, Germany
3 Center for Soft Nanoscience (SoN), Busso-Peus-Str. 10, 48149 Münster, Germany

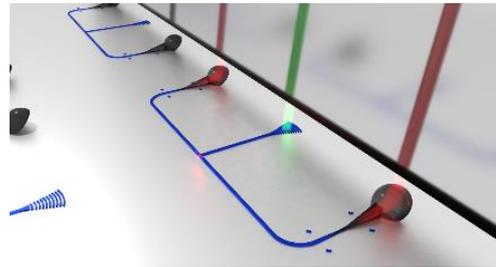

**ABSTRACT:** Solution processible colloidal quantum dots hold great promise for realizing single-photon sources embedded into scalable quantum technology platforms. However, the high-yield integration of large numbers of individually addressable colloidal quantum dots in a photonic circuit has remained an outstanding challenge. Here, we report on integrating individual colloidal core-shell quantum dots (CQDs) into a nanophotonic network that allows for excitation and efficient collection of single-photons via separate waveguide channels. An iterative electron beam lithography process provides a viable method to position single emitters at predefined positions on a chip with yield that approaches unity. Our work moves beyond the bulk optic paradigm of confocal microscopy and paves the way for supplying chip-scale quantum networks with single photons from large numbers of simultaneously controllable quantum emitters.

**Introduction** Nanoscale solid-state single-photon emitters are of central importance for experiments in quantum optics and realizations of quantum technologies[1]. Several emitter systems are currently being considered for such purposes, with epitaxial quantum dots showing leading performance regarding purity, indistinguishability, and stability[2–4], while presenting challenges in growth and integration with other quantum technology platforms. Colloidal quantum dots (CQDs)[5] also show great promise as single-photon sources, allowing tunable emission wavelengths, possibly even into the telecom range[6–8], through well controlled variations in geometry and material composition. Improvements in photon purity, indistinguishability and stability were recently achieved through advances in optimizing the shell composition and thickness[9–11] as well as smoothing the interface potential between core and shell[6,12] in CQD synthesis[13] or coupling to nanocavities[14]. Importantly, CQDs can be processed in solution therewith offering tremendous flexibility and scalability for integrating them with a broad range of material platforms and nanostructures[15–19].

To study systems aspects of complex quantum optical ensembles and realize practical applications of quantum technologies, large numbers of quantum emitters will have to be integrated into a network that allows for simultaneous optical addressing and manipulation. Photonic integrated circuits are well suited for this task because they provide reproducible replication of compact functional units, low channel loss and cost-efficient fabrication on monolithic chips. However, embedding large numbers of quantum emitters with high placement accuracy in a reconfigurable nanophotonic network as well as efficient interfaces between quantum emitters and low-loss waveguides remain challenging[20]. Moreover, high levels of intrinsic photoluminescence in the visible spectral range plague most established wide-band transparent waveguides made from high refractive index dielectrics, such as silicon nitride, which hampers experiments at the single-photon level. While extensive experimental efforts with single molecules[21–23], color centers in diamond[24–26], as well as self-assembled[27–29] and colloidal[18,30,31] quantum dots have been undertaken, photonic integrated circuits with large numbers of embedded individually addressable single photon sources have so far remained elusive.

Here we demonstrate single-photon emission from individual CQDs into nanophotonic circuits at room temperature and show how the approach can be scaled up to larger system size. We employ colloidal CdSeTe/ZnS core-shell quantum dots as single-photon sources that can be processed in solution, thus allowing for applying them in large numbers to lithographically patterned dies or even at the wafer-scale. We eliminate intrinsic material photoluminescence from dielectric waveguides upon optical excitation of CQDs by realizing nanophotonic circuits from tantalum pentoxide ($Ta_2O_5$) on insulator ($SiO_2$) thin films that benefit experiments with single photons through ultra-low fluorescent background in addition to low transmission loss (-1 dB/cm)[32]. Efficient and broadband interfaces between individual quantum emitters and nanophotonic circuits are achieved by positioning CQDs inside circular holes at the intersection of $Ta_2O_5$-waveguides for excitation and fluorescence collection, both of which only support a single transverse electric mode. Low loss and low fluorescent 3D optical interconnects between the waveguides and optical fibers further allow for efficiently extracting light from a chip and verifying the single-photon characteristics by recording the second-order autocorrelation function $g^{(2)}(\tau)$, which shows antibunching. By utilizing a multilayer lithography procedure with high overlay accuracy, we are able to iteratively fill vacant sites on a chip with CQDs while passivating occupied sites, thus providing a high-yield solution for integrating single-photon sources with nanophotonic circuits. Our approach paves the way for equipping scalable photonic integrated circuits with individually and simultaneously addressable single-photon sources. In combination with efficient optical interconnects such chips with massively parallelized integrated single-photon emitters will benefit applications in quantum communication [33], e.g. for outperforming laser-based quantum key distribution schemes[34].

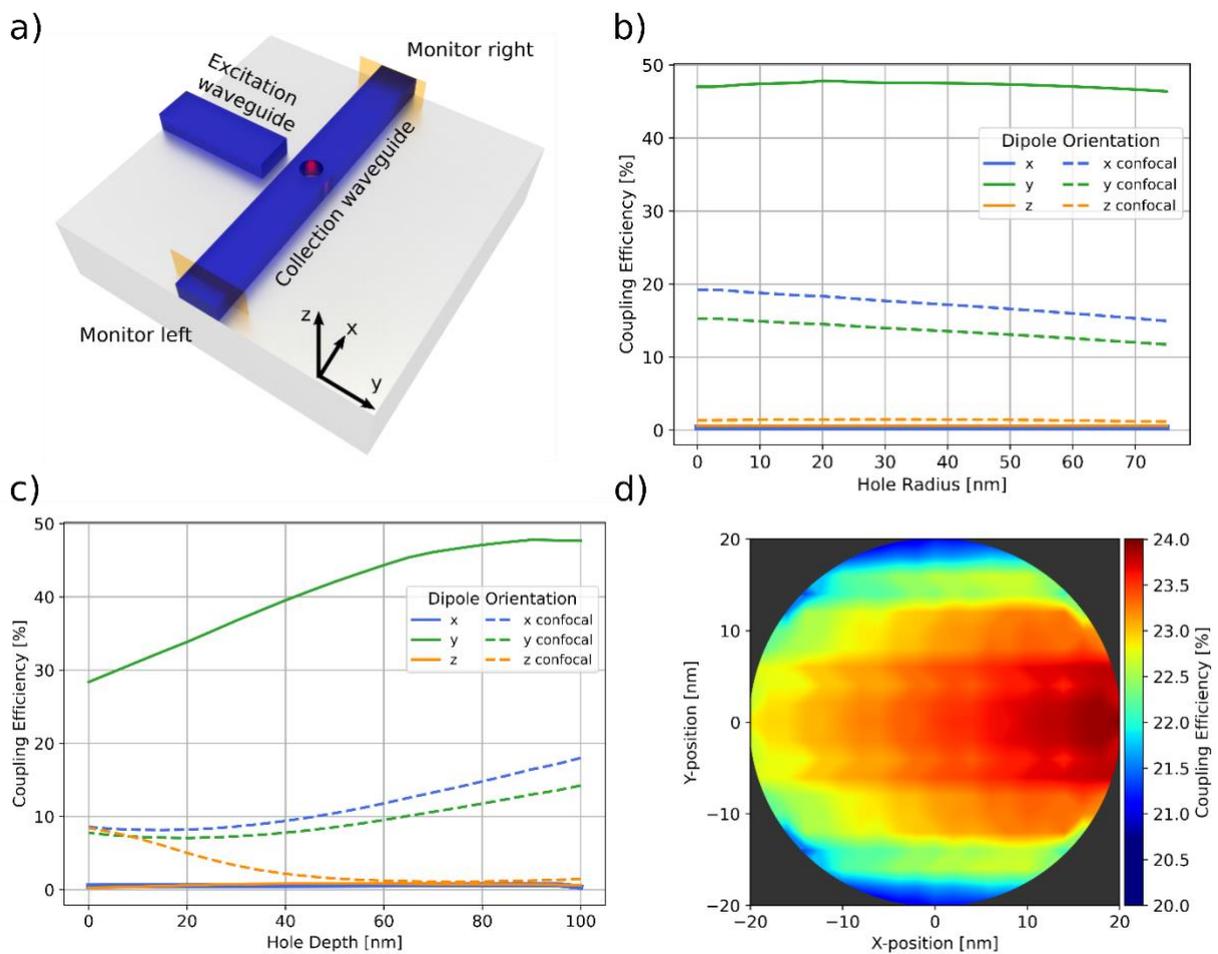

**Figure 1** FDTD numerical simulations of the waveguide crossing. a) Schematic illustration of the device layout considered for 3D numerical simulations. A dipole emitter representing a single CQD is placed inside the hole. At each side of the collection waveguide, monitors (orange planes) are positioned to determine the point vector flux inside of the waveguide. Additionally, another monitor is placed above the waveguide intersection, which represents the collection into an objective lens with 0.9 numerical aperture (see supplementary information). b) Coupling efficiency into the waveguide mode as a function of the waveguide hole radius (x/y/z-axis (blue, green, orange) oriented dipole – collected into the waveguide (solid) or the confocal microscope objective lens (dashed)). c) Coupling efficiency into the waveguide mode as a function of the hole depth in the waveguide. Dipole orientation along the x/y/z-axis (blue, green, orange) results in the corresponding collection efficiency into the waveguide (solid) or the confocal microscope objective lens (dashed). d) Coupling efficiency determined at right monitor in dependence of emitter position at the bottom of a 25 nm radius hole (100 nm depth) for a dipole oriented along the y-direction. Due to the symmetry of the simulated area, the result for the left monitor is mirrored vertically.

**CQD – waveguide coupling** We here consider the coupling of a CQD positioned at the intersection of two $Ta_2O_5$-waveguides, one for optical excitation with a 532 nm wavelength laser and one for fluorescence collection in the 650 – 750 nm spectral range, as shown in Fig. 1 a. While efficient optical excitation of the CQD can conveniently be achieved by setting appropriate laser power levels, the collection of photons will depend on positioning the quantum dot with respect to the waveguide mode. We perform finite-difference time-domain (FDTD) simulations for placing a single CQD within a hole inside the collection waveguide, as shown in Fig. 1 a. We systematically vary the hole radius, the hole depth and the position of the CQD within the hole to optimize the coupling conditions, as shown in Fig. 1 b – d. The CQD is here represented by a point dipole source emitting at a wavelength of 705 nm, embedded into spheres representing core and shell.

Firstly, we consider a CQD at the center of a 100 nm deep hole and calculate the optical power at two monitor planes extending somewhat beyond the 100 nm high, 700 nm wide collection waveguide, as shown in Fig. 1 a. We then derive the coupling efficiency into the waveguide from the sum of optical powers in both monitors normalized to the total emitted optical power. Upon varying the hole radius, we find a maximum overall coupling efficiency of 47% for a dipole emitter oriented along the y-direction and a 25 nm hole radius, as shown in Fig. 1 b. Remarkably, the coupling efficiency only changes marginally (45 % – 47 %) when varying the hole radius in the 0 – 75 nm range, highlighting the robustness of the geometry against fabrication imperfections. As expected, dipole emitters oriented along the x- and z-directions show no appreciable coupling into the guided optical mode. Moreover, we determine the collection efficiency into a 0.9 numerical aperture (NA) confocal microscope objective by placing a corresponding monitor plane above the waveguide intersection. Here, it is possible to collect light from dipole emitters oriented along x- and y-directions with maximal efficiencies of 19% and 15%, respectively, for hole radii below 10 nm. Larger hole radii reduce the collection efficiency in confocal microscopy down to 15% (x-direction) and 12% (y-direction), as shown in Fig. 1 b.

Secondly, we assess the influence of the hole depth, which could be controlled in a precisely timed etching process during device fabrication. In Fig. 1 c we consider a CQD at the center of a hole with a fixed radius of 25 nm and find the optimal coupling efficiency into the collection waveguide for hole depths in the 90 – 100 nm range, i.e. when etching (almost) through the entire waveguide (100 nm height), down to the buried oxide layer. Reducing the hole depth, the coupling efficiency gradually decrease to 28 % at 0 nm depth, i.e. the emitter lies on top of the waveguide. The preference for placing CQD closer to the bottom of the waveguide rather than the top reflects the higher refractive index of the substrate as compared to that of air surrounding the waveguide otherwise.

Lastly, we study the influence of the CQD position inside a 25 nm radius hole on the achievable coupling efficiency. Fig. 1 d shows the coupling efficiency when only considering one of the two power monitors, i.e. emission into one waveguide direction. A dipole emitter oriented along the y-direction thus yields coupling efficiencies in the 21 % - 24 % range, where the position in x-direction introduces a slight preference for emitting into one waveguide direction over another. Symmetry considerations imply a mirrored efficiency profile when calculating the power at the opposite monitor plane. The low overall dependence of coupling efficiency on CQD position shows that the coupling approach is robust against placement inaccuracies as long as the CQD ends up inside the hole. Control over the excitation and emission of a CQD's two orthogonal dipoles with respect to the waveguide directions could be achieved either by embedding the quantum dot structure into appropriately shaped nanoparticles, e.g. platelets[35], that align along preferred directions during on-chip assembly, or by modifying the nanophotonic access to the CQD through routing additional waveguides under various angles to each emitter site.

**Device design and fabrication** Our device layout is depicted schematically in Fig. 2 a, showing how 532 nm wavelength (green) light is supplied from an optical fiber array via the excitation waveguide (EWG) to the CQD inside the hole at the intersection with the collection waveguide (CWG), which in turn guides emitted photons (red) in opposite directions to two ports of the fiber array. For the latter we utilize 3D polymer structures produced by direct laser writing in polymer resist as fiber-chip interfaces because they offer efficient coupling over the entire spectral range of the CQD emission. For coupling the 532 nm laser from the fiber array into the excitation waveguide, however, we use grating couplers, which only transmit a restricted bandwidth and show negligible intrinsic material fluorescence under laser illumination, thus introducing minimal background in the 650 – 750 nm spectral range of interest.

We fabricate hundreds of $Ta_2O_5$ nanophotonic devices on a single chip and position solution-based CQDs at predefined locations by employing a lithographic technique that allows for an iterative procedure to increase the device yield as desired for scalable realizations of quantum technology. The fabrication steps of the $Ta_2O_5$ CQD devices are described in the supplementary information. From systematic parameter variation, we find an optimal compromise between single-CQD-per-site-yield and high simulated coupling efficiency for CQDs hosted in 35 nm radius holes in the collection waveguide. Figure 2 b) shows a scanning helium ion micrograph of the waveguide intersection of the final fabricated device, with a single CQD (white) located inside of the hole, as depicted in Fig. 2 c. Fig. 2 d shows a schematic illustration of a single CQD consisting of a CdSeTe core and ZnS shell, which are decorated with ligands on the outside.

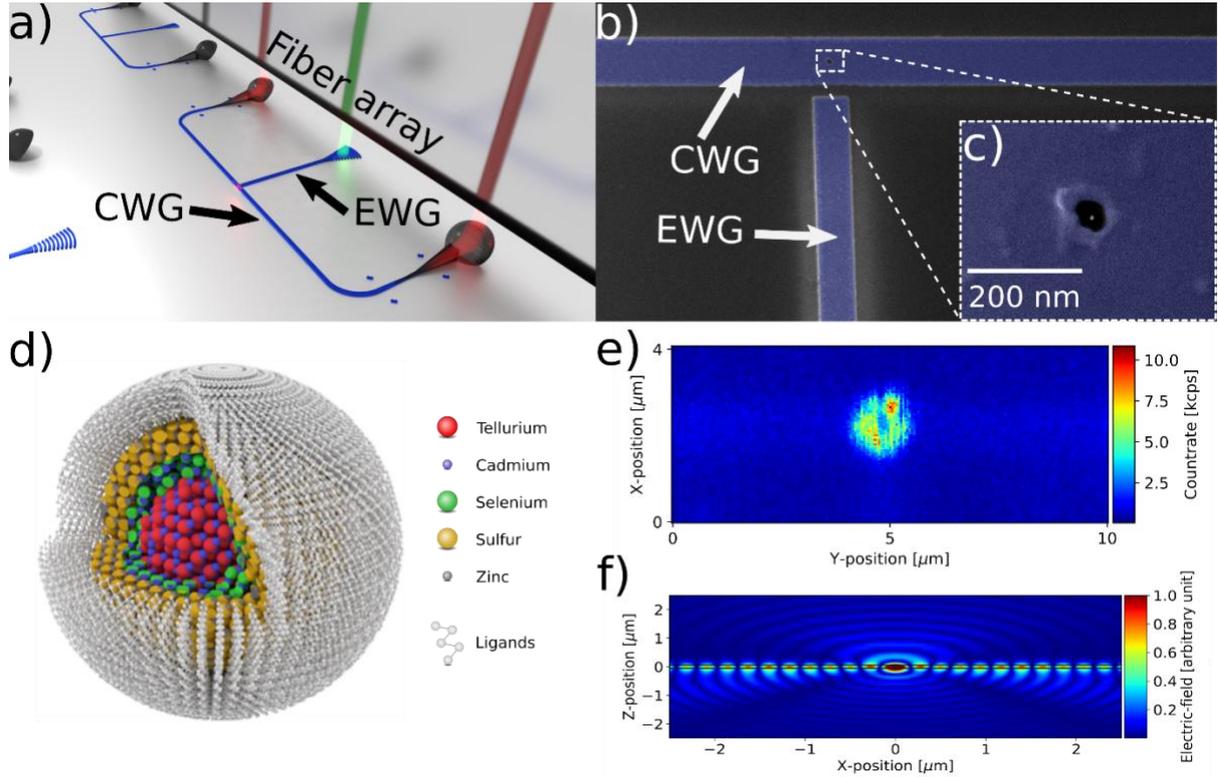

**Figure 2** a) Schematic illustration of the device layout of an emitter integrated photonic system. The waveguides are optically accessible from an optical fiber array via 3D polymer and grating coupler structures. A single colloidal quantum dot is located at the waveguide intersection of the collection (CWG) and excitation (EWG) waveguide. b) False color scanning helium ion micrograph (SHIM) of waveguide intersection, consisting of the excitation waveguide and the collection waveguide indicated in blue. c) Zoom in SHIM on the hole region in b). A single colloidal quantum dot (white) is located inside the center of the hole in the collection waveguide. d) Schematic illustration of single colloidal CdSeTe-core ZnS-shell quantum dot. e) Photoluminescence map of a single CQD located inside of a 35 nm radius hole at the intersection of $Ta_2O_5$ waveguides. The low intrinsic material fluorescence of the $Ta_2O_5$ waveguides is hardly discernable from the $SiO_2$ substrate. f) 2D FDTD simulation showing the propagation of light emitted from a CQD, which couples primarily into the waveguide mode.

To make single-photons emitted into nanophotonic waveguides available off-chip requires highly efficient optical fiber-chip interfaces. We achieve this by fabricating precisely aligned 3D coupling structures in direct laser writing of polymer resist. We here used a mixture of IP-Dip NPI (Nanoscribe GmbH), i.e. without photo initiator, and added a 2% weight fraction Irgacure 819 (Merck), which was shown to allow for fabricating polymer structures that exhibit low fluorescence levels upon optical illumination in the visible wavelength range[36]. Use of IP-Dip NPI ensures that the refractive index of the resist is matched to the objective of the direct laser writing tool and therefore allows for high-resolution dip-in lithography. The coupling structures employ an efficient and broadband total internal reflection design as previously described in[37], which was adjusted for the optical fibers used here[38].

**Device characterization and integrated measurements** Before characterizing the performance of the waveguide-integrated device, we assess the quantum emitter properties in confocal microscopy as a reference. We excite the CQDs located at the waveguide intersections with a 532 nm wavelength laser and collect the photoluminescence in the 650 – 750 nm spectral range using a 0.9 numerical aperture microscope objective. We record the photoluminescence while scanning the sample in lateral directions and observe the photon count rate behavior shown in Fig. 2 e. Due to blinking of the photoluminescence, characteristic for the CQDs used here[39,40], fluctuations of the intensity occur in the few ms up to few s time interval used for data collection at each scan position (pixel). The count rate fluctuations on short time scales give a first indication of the number of CQDs present at each site. Low levels of intrinsic fluorescence from the surrounding $Ta_2O_5$-on-insulator photonic structures[41,42] are mandatory for resolving such intensity fluctuations, with waveguides hardly discernable in Fig. 2 e. While the donut-shaped pattern of the photoluminescence collected from the CQD may be an indication of the emitter dipole orientation we cannot exclude it being an artifact originating from blinking. We further assess the statistical properties of the collected photoluminescence by recording the second-order autocorrelation function $g^{(2)}(\tau)$, utilizing a balanced beam splitter and single-photon counting modules in a Hanbury Brown and Twiss setup. A fit (orange) to the data (blue circles) reveals photon antibunching with $g^{(2)}(0)_{cm} = 0.32 \pm 0.08$ (see supplementary information), confirming occupation of the site with a CQD emitting single photons. The relatively large variations of the autocorrelation measurement are a consequence of the low signal levels recorded in confocal microscopy because light from the emitter is predominantly emitted into the waveguide mode, as shown in Fig. 2 f.

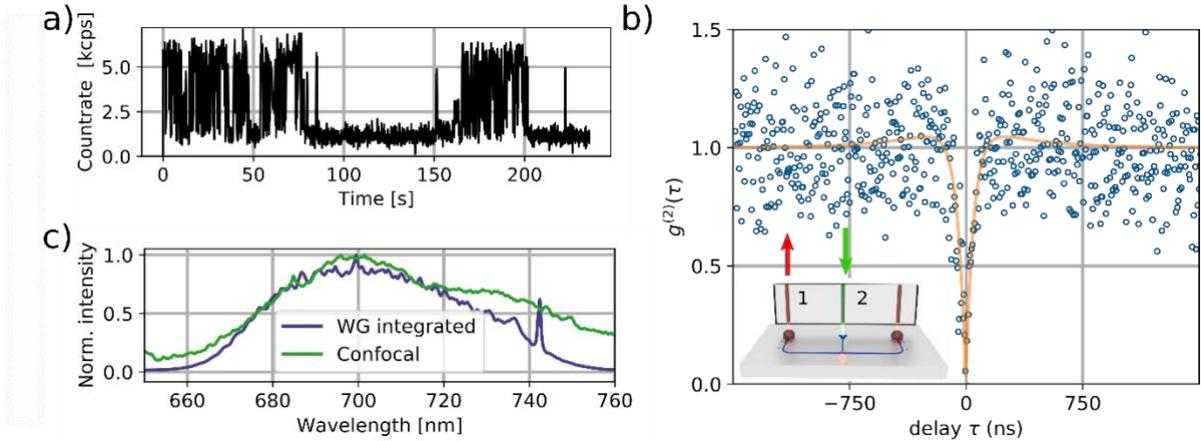

**Figure 3** Nanophotonic characterization of a CQD inside the hole at the intersection of excitation and collection waveguides a) Time trace of fluorescence intensity of a single CQD showing characteristic blinking behavior. b) Second-order autocorrelation function $g_{wg}^{(2)}(\tau)$ of waveguide-coupled CQD emission. The green arrow indicates the excitation path, the red arrow the collection path. A fit (orange) to the data (blue) yields $g_{wg}^{(2)}(0) = 0.04 \pm 0.08$. c) CQD photoluminescence spectrum recorded in confocal microscopy (green) and through a nanophotonic waveguide (WG, purple).

We then align the chip under an optical fiber array, which allows us to interface with integrated CQDs on up to 16 channels simultaneously, therewith overcoming a critical limitation of confocal microscopy setups that typically only allow access to one device at a time. To alleviate photo-stability issues that plague many CQDs[43,44], we perform our experiments at optical powers of approximately 1.5 µW inside the excitation waveguide, i.e. far below the saturation power. The count rate from an optically excited CQD recorded with single-photon avalanche diodes at one of the two collection ports reaches up to $(5521 \pm 98)$ counts/s, subject to the typical blinking behavior[39,45] of individual CQDs as shown in Fig. 3 a. Taking into account -3 dB attenuation from reading out only one of the two collection waveguide channels, -5.5 dB loss from the fiber-to-chip interface, -6.1 dB total transmission loss from spectral filters, and -1.5 dB loss in detection efficiency, we estimate the total emission rate of a single CQD into both directions of the collection waveguide as $2.3 \cdot 10^5$ photons/s at the modest pump power levels mentioned above. We note that for the spherical CQDs used here, the orientation of their dipoles with respect to the excitation and collection waveguides is uncontrolled and photobleaching gradually decreases the number of generated photons even at low excitation power, which both adds variability in the number of collected photons.

The second-order autocorrelation function $g^{(2)}(\tau)$ of light inside the collection waveguide is depicted in Fig. 3 b and shows background-corrected antibunching with $g_{wg}^{(2)}(0) = 0.04 \pm 0.08$, thus confirming the suitability of CQDs for supplying nanophotonic circuits with single-photons (see supplementary information for additional data). Here, the autocorrelation data was fitted with $g^{(2)}(\tau) = 1 + \frac{g_{func}^{(2)}(\tau)-1}{\rho^2}$, where $\rho = \frac{signal}{signal+background}$, and $g_{func}^{(2)}(\tau) = 1 - b \cdot e^{-\frac{|\tau|}{\tau_l}}$ applies for a two-level system, where b is a scaling parameter and $\tau_l$ is a time constant determined by the channel decay rate from the excited state at low excitation powers as we use in this experiment [46]. The background was determined from reference devices with similar design parameters but with no CQD present. Both the antibunching behavior and the resulting time constant $\tau_{l,wg} = (23.90 \pm 7.51)$ ns agree with our confocal characterization at a similar excitation power level and with literature values of similar CQDs [47].

We further utilize the second photoluminescence collection channel for analyzing the spectral composition of the light inside the waveguide, which is characteristic of the CQD emission, as illustrated in Fig. 3 c. For reference, we also show the spectrum recorded in confocal microscopy, with differences originating in the transmission characteristics of the nanophotonic waveguide and the 3D polymer coupling structures.

**Placement yield improvement** The integration of large numbers of quantum emitters with nanophotonic circuits requires a high-yield positioning process. State-of-the-art placement yields up to 40%[16] have been achieved with similar techniques to the one presented here by tuning aperture hole sizes and film thickness. While the drop-casting method inherently produces inhomogeneous films that result in yield limitations, methods utilizing

Langmuir-Blodgett[16,48] or capillary assembly[49] techniques produce more homogenous films and thus increase the yield somewhat. However, unity placement yield of single emitters has so far remained out of reach for all such approaches. Therefore, we here introduce an iterative procedure for selective CQD placement that allows approaching 100 % site occupancy.

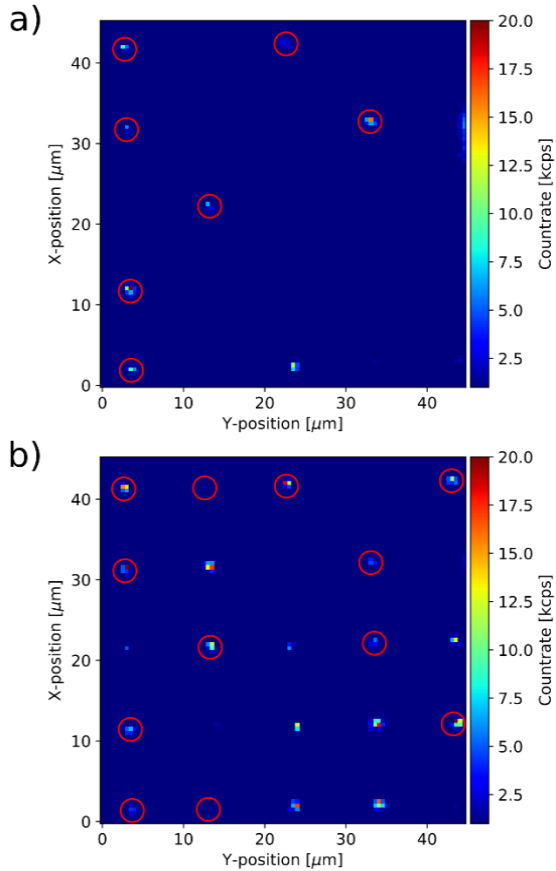

**Figure 4** Photoluminescence maps of 5×5 patterns of CQDs that were filled into 25 nm radius and 50 nm deep holes. a) After one iteration 8 out of 25 positions are occupied by fluorescent emitters. The red circles indicate sites that show photon statistics consistent with single emitter occupancy (7 out of 8). b) In a second iteration apertures were only patterned at vacant positions, resulting in a total of 20 out of 25 positions being occupied by fluorescent emitters. Red circles mark sites that show photon statistics consistent with single emitter occupancy (12 out of 20).

For our proof-of-principle demonstration, we utilize a thermally oxidized silicon chip that resembles our device substrates and pattern 5×5 hole arrays in PMMA with varying hole diameters via EBL. After development, we drop-cast a CQD solution onto the sample and remove the surplus of emitters in a lift-off process (see supplementary information). We then determine the placement yield in confocal microscopy under illumination with a 532 nm wavelength laser. Fig. 4 a shows the resulting photoluminescence map with 8 out of 25 sites (25 nm radius holes) occupied, where 7 of these show photon counting statistics consistent with single-emitter occupancy. We identify the locations of successfully placed emitters and eliminate the corresponding positions from our EBL mask pattern, such that a hole pattern with the remaining 17 vacant positions is used for the subsequent lithography step. Such selection of single-emitter sites avoids occupation with multiple emitters in later iterations. After performing EBL on PMMA, CQD-solution drop-casting and resist lift-off accordingly, we observe the photoluminescence map shown in Fig. 4 b. We find that all the emitters that were positioned during the first iteration and buried under PMMA in the second iteration stayed intact, and 12 additional sites were occupied during the second iteration. The total of 20 occupied sites out of 25 overall, of which 12 show single-photon emitter characteristics, correspond to a cumulative placement yield of 80 % (48 % single-photon emitters). For a mean CQD placement yield of 55% per iteration, on average only 6 iterations will be required to reach more than 99% site-occupation. The electron beam used for lithography can further be employed to efficiently neutralize sites occupied with multiple emitters during each iteration, such that unity single emitter placement yield becomes possible."

**Discussion and outlook** In conclusion, we have realized the first CQD-based single-photon source integrated with nanophotonic circuits. FDTD simulations showed that a straightforward design approach where a hole at a waveguide intersection hosts an individual CQD achieves 47 % coupling efficiency into guided modes over a broad bandwidth with excellent robustness against fabrication imperfections. At modest optical excitation powers, we launch approximately $2.3 \cdot 10^5$ photons/s in single-mode $Ta_2O_5$-on-insulator waveguides with very low intrinsic material fluorescence. We confirm the single-photon character of waveguide-coupled photoluminescence by observing antibunching behavior with $g_{wg}^{(2)}(0) = 0.04 \pm 0.08$, which agrees with corresponding confocal microscopy measurements. While the coherence properties of the CQDs used in our work are not yet suited for many quantum technology applications, improvements in colloidal quantum dot synthesis and optical control are on-going [5,10,12,50–52]. With such improvements or more advanced emitter systems of similar nanoparticle shapes our scalable approach, which readily adapts to operation in cryogenic environments, offers a highly promising perspective for realizing large numbers of integrated single-photon sources for quantum photonic application, such as high-rate quantum communication schemes[53,54]. Future work may further consider the manipulation of the CQD photo-physical properties, e.g. emission linewidth and rate, by embedding them into nanoscale optical resonators[14,24], thus benefitting more sophisticated applications in quantum technology. Moreover, our approach of addressing individual CQDs with nanophotonic waveguides, rather than bulky microscope setups, provides a novel method for investigating nanoscale quantum emitters in a compact fashion from a multitude of independent optical channels that can be operated in parallel.

Furthermore, we demonstrate a viable technique based on iterative lithographic patterning for approaching 100 % placement yield of CQDs on nanophotonic chips. In a proof-of-principle experiment, we achieved 80% site occupation after only two iterations. We anticipate that unity yield will be achievable with high probability after a small number of iterations when combining our technique with more elaborate placement procedures[16,48,49]. Moreover, we expect that our procedure straightforwardly adapts to other solution-processable quantum emitters, such as color centers in nanodiamonds[24], defects in silicon carbide[55], hexagonal boron nitride nanoparticles[56] or other nanoscale quantum systems[30].

## AUTHOR INFORMATION

**Corresponding Author**

* Carsten Schuck − orcid.org/ 0000-0002-9220-4021; Email: Carsten.schuck@uni-muenster.de

## FUNDING SOURCES

H.G. thanks the Studienstiftung des deutschen Volkes for financial support. C.S. acknowledges support from the Ministry for Culture and Science of North Rhine-Westphalia (421-8.03.03.02–130428). The authors acknowledge support by the German Research Foundation (DFG, CRC 1459).

## ACKNOWLEDGMENT

We would like to thank the Münster Nanofabrication Facility (MNF) for their support in nanofabrication related matters.

## REFERENCES

(1) Aharonovich, I.; Englund, D.; Toth, M. Solid-State Single-Photon Emitters. *Nat. Photonics* **2016**, *10* (10), 631–641. https://doi.org/10.1038/nphoton.2016.186.

(2) Somaschi, N.; Giesz, V.; De Santis, L.; Loredo, J. C.; Almeida, M. P.; Hornecker, G.; Portalupi, S. L.; Grange, T.; Antón, C.; Demory, J.; Gómez, C.; Sagnes, I.; Lanzillotti-Kimura, N. D.; Lemaítre, A.; Auffeves, A.; White, A. G.; Lanco, L.; Senellart, P. Near-Optimal Single-Photon Sources in the Solid State. *Nat. Photonics* **2016**, *10* (5), 340–345. https://doi.org/10.1038/nphoton.2016.23.

(3) Kuhlmann, A. V.; Prechtel, J. H.; Houel, J.; Ludwig, A.; Reuter, D.; Wieck, A. D.; Warburton, R. J. Transform-Limited Single Photons from a Single Quantum Dot. *Nat. Commun.* **2015**, *6* (1), 8204. https://doi.org/10.1038/ncomms9204.

(4) Ding, X.; He, Y.; Duan, Z.-C.; Gregersen, N.; Chen, M.-C.; Unsleber, S.; Maier, S.; Schneider, C.; Kamp, M.; Höfling, S.; Lu, C.-Y.; Pan, J.-W. On-Demand Single Photons with High Extraction Efficiency and Near-Unity Indistinguishability from a Resonantly Driven Quantum Dot in a Micropillar. *Phys. Rev. Lett.* **2016**, *116* (2), 020401. https://doi.org/10.1103/PhysRevLett.116.020401.

(5) Kagan, C. R.; Bassett, L. C.; Murray, C. B.; Thompson, S. M. Colloidal Quantum Dots as Platforms for Quantum Information Science. *Chem. Rev.* **2021**, *121* (5), 3186–3233. https://doi.org/10.1021/acs.chemrev.0c00831.

(6) Krishnamurthy, S.; Singh, A.; Hu, Z.; Blake, A. V.; Kim, Y.; Singh, A.; Dolgopolova, E. A.; Williams, D. J.; Piryatinski, A.; Malko, A. V.; Htoon, H.; Sykora, M.; Hollingsworth, J. A. PbS/CdS Quantum Dot Room-Temperature Single-Emitter Spectroscopy Reaches the Telecom O and S Bands via an Engineered Stability. *ACS Nano* **2021**, *15* (1), 575–587. https://doi.org/10.1021/acsnano.0c05907.

(7) Bisschop, S.; Geiregat, P.; Aubert, T.; Hens, Z. The Impact of Core/Shell Sizes on the Optical Gain Characteristics of CdSe/CdS Quantum Dots. *ACS Nano* **2018**, *12* (9), 9011–9021. https://doi.org/10.1021/acsnano.8b02493.

(8) Lu, H.; Carroll, G. M.; Neale, N. R.; Beard, M. C. Infrared Quantum Dots: Progress, Challenges, and Opportunities. *ACS Nano* **2019**, *13* (2), 939–953. https://doi.org/10.1021/acsnano.8b09815.

(9) Ji, B.; Koley, S.; Slobodkin, I.; Remennik, S.; Banin, U. ZnSe/ZnS Core/Shell Quantum Dots with Superior Optical Properties through Thermodynamic Shell Growth. *Nano Lett.* **2020**, *20* (4), 2387–2395. https://doi.org/10.1021/acs.nanolett.9b05020.

(10) Htoon, H.; Malko, A. V.; Bussian, D.; Vela, J.; Chen, Y.; Hollingsworth, J. A.; Klimov, V. I. Highly Emissive Multiexcitons in Steady-State Photoluminescence of Individual "Giant" CdSe/CdS Core/Shell Nanocrystals. *Nano Lett.* **2010**, *10* (7), 2401–2407. https://doi.org/10.1021/nl1004652.

(11) Zhou, J.; Zhu, M.; Meng, R.; Qin, H.; Peng, X. Ideal CdSe/CdS Core/Shell Nanocrystals Enabled by Entropic Ligands and Their Core Size-, Shell Thickness-, and Ligand-Dependent Photoluminescence Properties. *J. Am. Chem. Soc.* **2017**, *139* (46), 16556–16567. https://doi.org/10.1021/jacs.7b07434.

(12) Climente, J. I.; Movilla, J. L.; Planelles, J. Auger Recombination Suppression in Nanocrystals with Asymmetric Electron-Hole Confinement. *Small* **2012**, *8* (5), 754–759. https://doi.org/10.1002/smll.201101740.

(13) Pu, Y.; Cai, F.; Wang, D.; Wang, J.-X.; Chen, J.-F. Colloidal Synthesis of Semiconductor Quantum Dots toward Large-Scale Production: A Review. *Ind. Eng. Chem. Res.* **2018**, *57* (6), 1790–1802. https://doi.org/10.1021/acs.iecr.7b04836.

(14) Saxena, A.; Chen, Y.; Ryou, A.; Sevilla, C. G.; Xu, P.; Majumdar, A. Improving Indistinguishability of Single Photons from Colloidal Quantum Dots Using Nanocavities. *ACS Photonics* **2019**, *6* (12), 3166–3173. https://doi.org/10.1021/acsphotonics.9b01481.

(15) Manfrinato, V. R.; Wanger, D. D.; Strasfeld, D. B.; Han, H.-S.; Marsili, F.; Arrieta, J. P.; Mentzel, T. S.; Bawendi, M. G.; Berggren, K. K. Controlled Placement of Colloidal Quantum Dots in Sub-15 Nm Clusters. *Nanotechnology* **2013**, *24* (12), 125302. https://doi.org/10.1088/0957-4484/24/12/125302.

(16) Xie, W.; Gomes, R.; Aubert, T.; Bisschop, S.; Zhu, Y.; Hens, Z.; Brainis, E.; Van Thourhout, D. Nanoscale and Single-Dot Patterning of Colloidal Quantum Dots. *Nano Lett.* **2015**, *15* (11), 7481–7487. https://doi.org/10.1021/acs.nanolett.5b03068.

(17) Zhang, Q.; Dang, C.; Urabe, H.; Wang, J.; Sun, S.; Nurmikko, A. Large Ordered Arrays of Single Photon Sources Based on II–VI Semiconductor Colloidal Quantum Dot. *Opt. Express* **2008**, *16* (24), 19592. https://doi.org/10.1364/OE.16.019592.

(18) Chen, Y.; Ryou, A.; Friedfeld, M. R.; Fryett, T.; Whitehead, J.; Cossairt, B. M.; Majumdar, A. Deterministic Positioning of Colloidal Quantum Dots on Silicon Nitride Nanobeam Cavities. *Nano Lett.* **2018**, *18* (10), 6404–6410. https://doi.org/10.1021/acs.nanolett.8b02764.

(19) Chen, J.; Rong, K. Nanophotonic Devices and Circuits Based on Colloidal Quantum Dots. *Mater. Chem. Front.* **2021**, *5* (12), 4502–4537. https://doi.org/10.1039/D0QM01118E.

(20) Lee, J.; Leong, V.; Kalashnikov, D.; Dai, J.; Gandhi, A.; Krivitsky, L. A. Integrated Single Photon Emitters. *AVS Quantum Sci.* **2020**, *2* (3), 031701. https://doi.org/10.1116/5.0011316.

(21) Lombardi, P.; Ovvyan, A. P.; Pazzagli, S.; Mazzamuto, G.; Kewes, G.; Neitzke, O.; Gruhler, N.; Benson, O.; Pernice, W. H. P.; Cataliotti, F. S.; Toninelli, C. Photostable Molecules on Chip: Integrated Sources of Nonclassical Light. *ACS Photonics* **2018**, *5* (1), 126–132. https://doi.org/10.1021/acsphotonics.7b00521.

(22) Türschmann, P.; Rotenberg, N.; Renger, J.; Harder, I.; Lohse, O.; Utikal, T.; Götzinger, S.; Sandoghdar, V. Chip-Based All-Optical Control of Single Molecules Coherently Coupled to a Nanoguide. *Nano Lett.* **2017**, *17* (8), 4941–4945. https://doi.org/10.1021/acs.nanolett.7b02033.

(23) Rattenbacher, D.; Shkarin, A.; Renger, J.; Utikal, T.; Götzinger, S.; Sandoghdar, V. Coherent Coupling of Single Molecules to On-Chip Ring Resonators. *New J. Phys.* **2019**, *21* (6), 062002. https://doi.org/10.1088/1367-2630/ab28b2.

(24) Schrinner, P. P. J.; Olthaus, J.; Reiter, D. E.; Schuck, C. Integration of Diamond-Based Quantum Emitters with Nanophotonic Circuits. *Nano Lett.* **2020**, *20* (11), 8170–8177. https://doi.org/10.1021/acs.nanolett.0c03262.

(25) Fehler, K. G.; Ovvyan, A. P.; Antoniuk, L.; Lettner, N.; Gruhler, N.; Davydov, V. A.; Agafonov, V. N.; Pernice, W. H. P.; Kubanek, A. Purcell-Enhanced Emission from Individual SiV−Center in Nanodiamonds Coupled to a Si$_3$N$_4$-Based, Photonic Crystal Cavity. *Nanophotonics* **2020**, *9* (11), 3655–3662. https://doi.org/10.1515/nanoph-2020-0257.

(26) Wan, N. H.; Lu, T. J.; Chen, K. C.; Walsh, M. P.; Trusheim, M. E.; De Santis, L.; Bersin, E. A.; Harris, I. B.; Mouradian, S. L.; Christen, I. R.; Bielejec, E. S.; Englund, D. Large-Scale Integration of Artificial Atoms in Hybrid Photonic Circuits. *Nature* **2020**, *583* (7815), 226–231. https://doi.org/10.1038/s41586-020-2441-3.

(27) Laurent, S.; Varoutsis, S.; Le Gratiet, L.; Lemaître, A.; Sagnes, I.; Raineri, F.; Levenson, A.; Robert-Philip, I.; Abram, I. Indistinguishable Single Photons from a Single-Quantum Dot in a Two-Dimensional Photonic Crystal Cavity. *Appl. Phys. Lett.* **2005**, *87* (16), 163107. https://doi.org/10.1063/1.2103397.

(28) Zadeh, I. E.; Elshaari, A. W.; Jöns, K. D.; Fognini, A.; Dalacu, D.; Poole, P. J.; Reimer, M. E.; Zwiller, V. Deterministic Integration of Single Photon Sources in Silicon Based Photonic Circuits. *Nano Lett.* **2016**, *16* (4), 2289–2294. https://doi.org/10.1021/acs.nanolett.5b04709.

(29) Schnauber, P.; Singh, A.; Schall, J.; Park, S. I.; Song, J. D.; Rodt, S.; Srinivasan, K.; Reitzenstein, S.; Davanco, M. Indistinguishable Photons from Deterministically Integrated Single Quantum Dots in Heterogeneous GaAs/Si$_3$N$_4$ Quantum Photonic Circuits. *Nano Lett.* **2019**, *19* (10), 7164–7172. https://doi.org/10.1021/acs.nanolett.9b02758.

(30) Elsinger, L.; Gourgues, R.; Zadeh, I. E.; Maes, J.; Guardiani, A.; Bulgarini, G.; Pereira, S. F.; Dorenbos, S. N.; Zwiller, V.; Hens, Z.; Van Thourhout, D. Integration of Colloidal PbS/CdS Quantum Dots with Plasmonic Antennas and Superconducting Detectors on a Silicon Nitride Photonic Platform. *Nano Lett.* **2019**, *19* (8), 5452–5458. https://doi.org/10.1021/acs.nanolett.9b01948.

(31) Weeber, J.-C.; Hammani, K.; Colas-des-Francs, G.; Bouhelier, A.; Arocas, J.; Kumar, A.; Eloi, F.; Buil, S.; Quélin, X.; Hermier, J.-P.; Nasilowski, M.; Dubertret, B. Colloidal Quantum Dot Integrated Light Sources for Plasmon Mediated Photonic Waveguide Excitation. *ACS Photonics* **2016**, *3* (5), 844–852. https://doi.org/10.1021/acsphotonics.6b00054.

(32) Splitthoff, L.; Wolff, M. A.; Grottke, T.; Schuck, C. Tantalum Pentoxide Nanophotonic Circuits for Integrated Quantum Technology. *Opt. Express* **2020**, *28* (8), 11921. https://doi.org/10.1364/oe.388080.

(33) Gisin, N.; Thew, R. Quantum Communication. *Nat. Photonics* **2007**, *1* (3), 165–171. https://doi.org/10.1038/nphoton.2007.22.

(34) Takemoto, K.; Nambu, Y.; Miyazawa, T.; Sakuma, Y.; Yamamoto, T.; Yorozu, S.; Arakawa, Y. Quantum Key Distribution over 120 Km Using Ultrahigh Purity Single-Photon Source and Superconducting Single-Photon Detectors. *Sci. Rep.* **2015**, *5* (1), 14383. https://doi.org/10.1038/srep14383.

(35) Feng, F.; Nguyen, L. T.; Nasilowski, M.; Nadal, B.; Dubertret, B.; Maître, A.; Coolen, L. Probing the Fluorescence Dipoles of Single Cubic CdSe/CdS Nanoplatelets with Vertical or Horizontal Orientations. *ACS Photonics* **2018**, *5* (5), 1994–1999. https://doi.org/10.1021/acsphotonics.7b01475.

(36) Shi, Q.; Sontheimer, B.; Nikolay, N.; Schell, A. W.; Fischer, J.; Naber, A.; Benson, O.; Wegener, M. Wiring up Pre-Characterized Single-Photon Emitters by Laser Lithography. *Sci. Rep.* **2016**, *6* (1), 1–7. https://doi.org/10.1038/srep31135.

(37) Gehring, H.; Eich, A.; Schuck, C.; Pernice, W. H. P. Broadband Out-of-Plane Coupling at Visible Wavelengths. *Opt. Lett.* **2019**, *44* (20), 5089–5092. https://doi.org/10.1364/OL.44.005089.

(38) Gehring, H.; Blaicher, M.; Eich, A.; Hartmann, W.; Varytis, P.; Busch, K.; Schuck, C.; Wegener, M.; Pernice, W. H. P. Broadband Fiber-to-Chip Coupling in Different Wavelength Regimes Realized by 3D-Structures. In *Conference on Lasers and Electro-Optics*; OSA: Washington, D.C., 2020; Vol. 2020-May, pp 3–4. https://doi.org/10.1364/CLEO_AT.2020.JTh2B.22.

(39) Guo, W.; Tang, J.; Zhang, G.; Li, B.; Yang, C.; Chen, R.; Qin, C.; Hu, J.; Zhong, H.; Xiao, L.; Jia, S. Photoluminescence Blinking and Biexciton Auger Recombination in Single Colloidal Quantum Dots with Sharp and Smooth Core/Shell Interfaces. *J. Phys. Chem. Lett.* **2021**, *12* (1), 405–412. https://doi.org/10.1021/acs.jpclett.0c03065.

(40) Yuan, G.; Gómez, D. E.; Kirkwood, N.; Boldt, K.; Mulvaney, P. Two Mechanisms Determine Quantum Dot Blinking. *ACS Nano* **2018**, *12* (4), 3397–3405. https://doi.org/10.1021/acsnano.7b09052.

(41) Zhao, Y.; Jenkins, M.; Measor, P.; Leake, K.; Liu, S.; Schmidt, H.; Hawkins, A. R. Hollow Waveguides with Low Intrinsic Photoluminescence Fabricated with Ta2O5 and SiO2 Films. *Appl. Phys. Lett.* **2011**, *98* (9), 091104. https://doi.org/10.1063/1.3561749.

(42) Kaji, T.; Yamada, T.; Ueda, R.; Xu, X.; Otomo, A. Fabrication of Two-Dimensional Ta2O5 Photonic Crystal Slabs with Ultra-Low Background Emission toward Highly Sensitive Fluorescence Spectroscopy. *Opt. Express* **2011**, *19* (2), 1422–1428. https://doi.org/10.1364/OE.19.001422.

(43) Brokmann, X.; Hermier, J.-P.; Messin, G.; Desbiolles, P.; Bouchaud, J.-P.; Dahan, M. Statistical Aging and Nonergodicity in the Fluorescence of Single Nanocrystals. *Phys. Rev. Lett.* **2003**, *90* (12), 120601. https://doi.org/10.1103/PhysRevLett.90.120601.

(44) Cheng, C.-Y.; Mao, M.-H. Photo-Stability and Time-Resolved Photoluminescence Study of Colloidal CdSe/ZnS Quantum Dots Passivated in Al 2 O 3 Using Atomic Layer Deposition. *J. Appl. Phys.* **2016**, *120* (8), 083103. https://doi.org/10.1063/1.4961425.

(45) Lee, S. F.; Osborne, M. A. Brightening, Blinking, Bluing and Bleaching in the Life of a Quantum Dot: Friend or Foe? *ChemPhysChem*. 2009, pp 2174–2191. https://doi.org/10.1002/cphc.200900200.

(46) Messin, G.; Hermier, J. P.; Giacobino, E.; Desbiolles, P.; Dahan, M. Bunching and Antibunching in the Fluorescence of Semiconductor Nanocrystals. *Opt. Lett.* **2001**, *26* (23), 1891. https://doi.org/10.1364/ol.26.001891.

(47) Yadav, R. K.; Liu, W.; Li, R.; Odom, T. W.; Agarwal, G. S.; Basu, J. K. Room-Temperature Coupling of Single Photon Emitting Quantum Dots to Localized and Delocalized Modes in a Plasmonic Nanocavity Array. *ACS Photonics* **2021**, *8* (2), 576–584. https://doi.org/10.1021/acsphotonics.0c01635.

(48) Lambert, K.; Moreels, I.; Van Thourhout, D.; Hens, Z. Quantum Dot Micropatterning on Si. *Langmuir* **2008**, *24* (11), 5961–5966. https://doi.org/10.1021/la703664r.

(49) Flauraud, V.; Mastrangeli, M.; Bernasconi, G. D.; Butet, J.; Alexander, D. T. L.; Shahrabi, E.; Martin, O. J. F.; Brugger, J. Nanoscale Topographical Control of Capillary Assembly of Nanoparticles. *Nat. Nanotechnol.* **2017**, *12* (1), 73–80. https://doi.org/10.1038/nnano.2016.179.

(50) Chen, Y.; Vela, J.; Htoon, H.; Casson, J. L.; Werder, D. J.; Bussian, D. A.; Klimov, V. I.; Hollingsworth, J. A. "Giant" Multishell CdSe Nanocrystal Quantum Dots with Suppressed Blinking. *J. Am. Chem. Soc.* **2008**, *130* (15), 5026–5027. https://doi.org/10.1021/ja711379k.

(51) Mahler, B.; Spinicelli, P.; Buil, S.; Quelin, X.; Hermier, J. P.; Dubertret, B. Towards Non-Blinking Colloidal Quantum Dots. *Nat. Mater.* **2008**, *7* (8), 659–664. https://doi.org/10.1038/nmat2222.

(52) Shi, J.; Sun, W.; Utzat, H.; Farahvash, A.; Gao, F. Y.; Zhang, Z.; Barotov, U.; Willard, A. P.; Nelson, K. A.; Bawendi, M. G. All-Optical Fluorescence Blinking Control in Quantum Dots with Ultrafast Mid-Infrared Pulses. *Nat. Nanotechnol.* **2021**, 1–7. https://doi.org/10.1038/s41565-021-01016-w.

(53) Lo, H. K.; Curty, M.; Tamaki, K. Secure Quantum Key Distribution. *Nature Photonics*. 2014, pp 595–604. https://doi.org/10.1038/nphoton.2014.149.

(54) Scarani, V.; Bechmann-Pasquinucci, H.; Cerf, N. J.; Dušek, M.; Lütkenhaus, N.; Peev, M. The Security of Practical Quantum Key Distribution. *Rev. Mod. Phys.* **2009**, *81* (3), 1301–1350. https://doi.org/10.1103/RevModPhys.81.1301.

(55) Lukin, D. M.; Guidry, M. A.; Vučković, J. Integrated Quantum Photonics with Silicon Carbide: Challenges and Prospects. *PRX Quantum* **2020**, *1* (2). https://doi.org/10.1103/prxquantum.1.020102.

(56) Duong, N. M. H.; Glushkov, E.; Chernev, A.; Navikas, V.; Comtet, J.; Nguyen, M. A. P.; Toth, M.; Radenovic, A.; Tran, T. T.; Aharonovich, I. Facile Production of Hexagonal Boron Nitride Nanoparticles by Cryogenic Exfoliation. *Nano Lett.* **2019**, *19* (8), 5417–5422. https://doi.org/10.1021/acs.nanolett.9b01913.

# Supporting information:

# Single photon emission from individual nanophotonic-integrated colloidal quantum dots

*Alexander Eich[1,2,3], Tobias C. Spiekermann[1,2,3], Helge Gehring[1,2,3], Lisa Sommer[1,2,3], Julian R. Bankwitz[1,2,3], Philip P.J. Schrinner[1,2,3], Johann A. Preuß[1,2], Steffen Michaelis de Vasconcellos[1,2], Rudolf Bratschitsch[1,2], Wolfram H. P. Pernice[1,2,3], Carsten Schuck[1,2,3], \**

1 Institute of Physics, University of Münster, Wilhelm-Klemm-Str. 10, 48149 Münster, Germany
2 Center for Nanotechnology (CeNTech), Heisenbergstr. 11, 48149 Münster, Germany
3 Center for Soft Nanoscience (SoN), Busso-Peus-Str. 10, 48149 Münster, Germany

E-mail: Carsten.schuck@uni-muenster.de

**Finite-difference time-domain simulation of the CQD – waveguide coupling**

To design the colloidal quantum dot (CQD) – nanophotonic waveguide interface, we performed finite-difference time-domain (FDTD) simulations of the region of interest using the Lumerical® software package. The goal was to find a geometry, which shows robustness against fabrication and placement deviations, and which is optimized regarding the coupling efficiency of the emission of a single CQD into the supported waveguide mode of $Ta_2O_5$ waveguides. Figure S1 shows the simulation setup of the waveguide intersection (S1 a) side view, S1 b) top view).  The single CQD is represented by a dipole source placed inside a sphere consisting of the core-shell material composition of the CQDs used here, which is placed inside the hole of the horizontal waveguide. Detector planes on each end of the waveguide observe the emission into the waveguide mode. The detector planes reach outside of the waveguide geometries to account for the evanescent field of the optical mode.  Additionally, further detector planes were placed above and below the waveguide intersection region. The plane above the intersection serves for estimating the collection efficiency that can be achieved in confocal microscopy with 0.9 numerical aperture. The detector plane placed in the substrate below the intersection region allows for estimating optical loss into the substrate. In Table S1 the sum of the light emitted into all the detection planes is shown for three orthogonal dipole orientations. For the preferred polarization (y direction) orientation we find over 80% of the emitted light, primarily coupled into the waveguide mode. For the other polarization orientations, the detector planes register >50% (x direction) and >40%

**Table S1** Total collected light for the sum of the detector planes for the different dipole source orientations and hole radii.

| Hole Radius [nm] | Total x orientation | Total y orientation | Total z orientation |
| --- | --- | --- | --- |
| 40 | 51.3% | 83.3% | 40.3% |
| 37.5 | 51.5% | 83.4% | 40.3% |
| 35 | 51.6% | 83.5% | 40.3% |
| 32.5 | 51.8% | 83.6% | 40.3% |
| 30 | 51.9% | 83.7% | 40.3% |
| 27.5 | 52.2% | 83.9% | 40.3% |
| 25 | 52.4% | 84.1% | 40.3% |
| 22.5 | 52.7% | 84.4% | 40.3% |
| 20 | 53.0% | 84.6% | 40.3% |

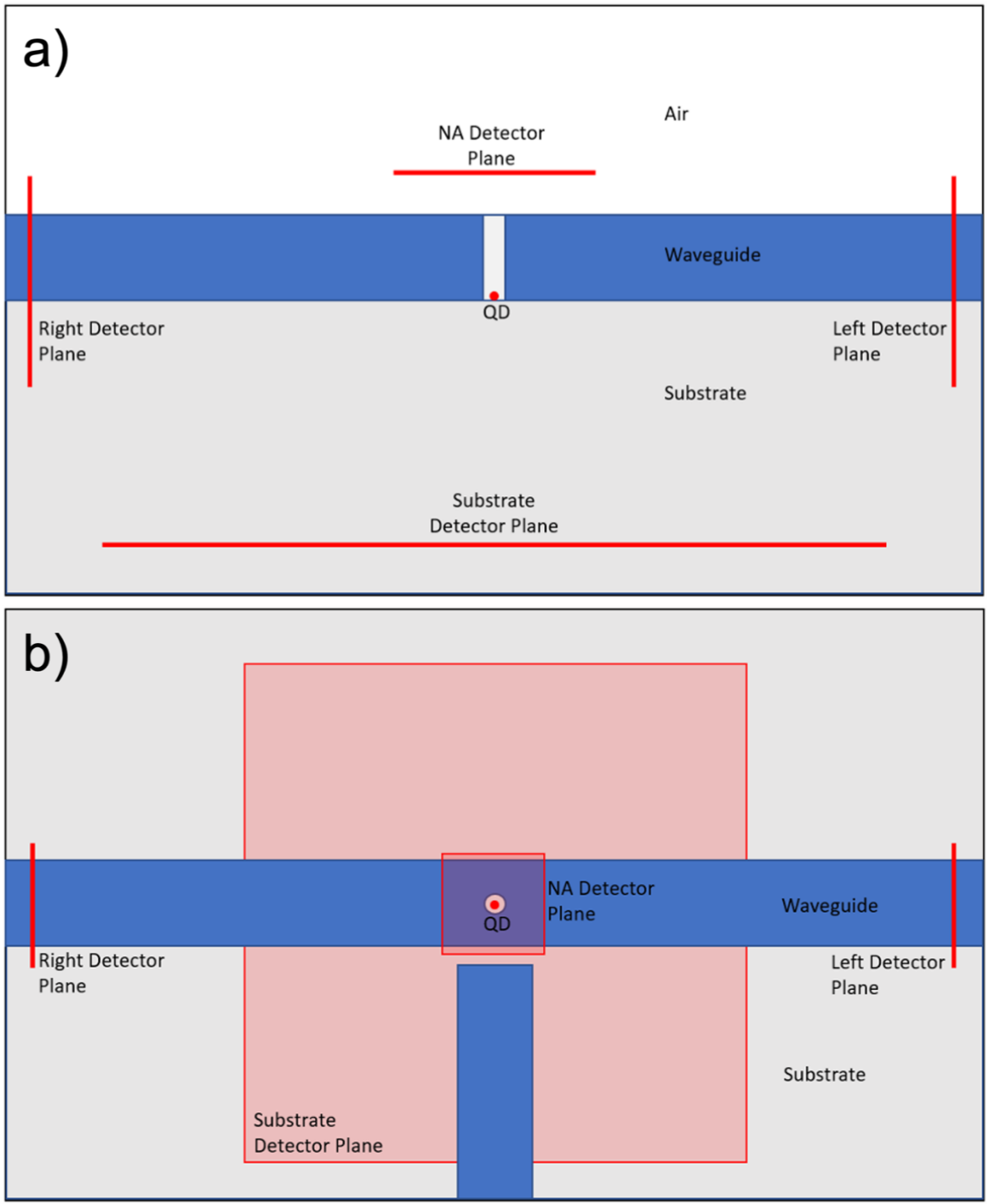

**Figure S1** Schematic illustration of the FDTD simulation setup consisting in the waveguide geometries (blue), the substrate (grey), the detector planes and the CQD (red). a) Side view of the intersection, b) top view of the intersection.

(z direction) of the emission, which indicates that a large fraction of the emitted light propagates into directions not covered by detector planes, as expected for such orientations of the emitting dipole. For a dipole oriented along the y direction the propagating optical fields are depicted in Figure S2. In Figure S2 a) the x-z-plane is displayed and shows that the intensity is predominantly confined inside the waveguide mode. However, a nonnegligible fraction of the emission is scattered into the substrate and into the confocal detector plane. Furthermore, when considering the x-y-plane, as shown in Figure S2 b), the predominant confinement of emitted light inside the waveguide mode is confirmed and reveals minor optical contributions dissipated into the x-y-plane. We conclude that most of the light, which is not confined inside the waveguide mode is scattered into the substrate.

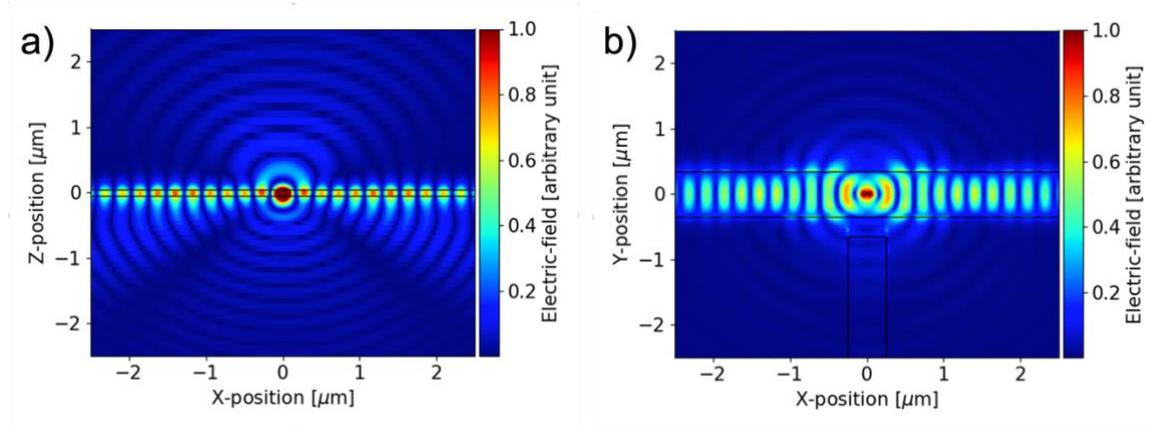

**Figure S2** 2D FDTD simulation showing the propagation of light emitted from a CQD with a dipole oriented in y-direction at the waveguide intersection region. a) shows the x-z-plane and b) shows the x-y-plane.

**Integration of colloidal quantum dots with tantalum pentoxide waveguides**

The fabrication of the device requires precisely aligned lithography steps to guarantee high overlay accuracy between the (i) nanophotonic circuits, (ii) the CQDs, and (iii) the 3D polymer couplers. We hence first produced alignment markers by physical vapor deposition in a lift-off process on a 100 nm $Ta_2O_5$ thin film that was sputter-deposited on top of a 2.6 µm thick silicon dioxide layer on a silicon substrate. We then fabricated hundreds of aligned nanophotonic devices in electron beam lithography (EBL) and reactive ion etching of the $Ta_2O_5$ thin film. We chose to pattern the holes at the waveguide intersections along with all other nanophotonic devices, implying hole depths of 100 nm. This procedure avoids an additional lithography step at the expense of very minor deviations from optimal CQD-waveguide coupling performance (see Fig. 1 c in the main text). In Figure S3, step 1, the resulting collection waveguide is displayed schematically. For the subsequent positioning of CQD inside the holes, we spin-coated the chip with a 135 nm thin Poly(methyl methacrylate) (PMMA) layer (Fig S3, step 2) and employed EBL-patterning for producing accurately aligned apertures with 50 nm diameter at each hole position (Fig S3, step 3). We filled the holes by drop-casting a decane-solution of CQDs (Invitrogen™ Qdot™ 705 ITK™ Organic Quantum Dots) in a saturated acetone atmosphere to slow the drying process (Fig S3, step 4). Afterwards, we removed the PMMA layer as well as the surplus of colloidal quantum dots via a lift-off procedure (Fig S3, step 5). In our parameter variation, we find an optimal compromise between single-CQD-per-site-yield and high simulated coupling efficiency for CQDs hosted in 35 nm radius holes in the collection waveguide.

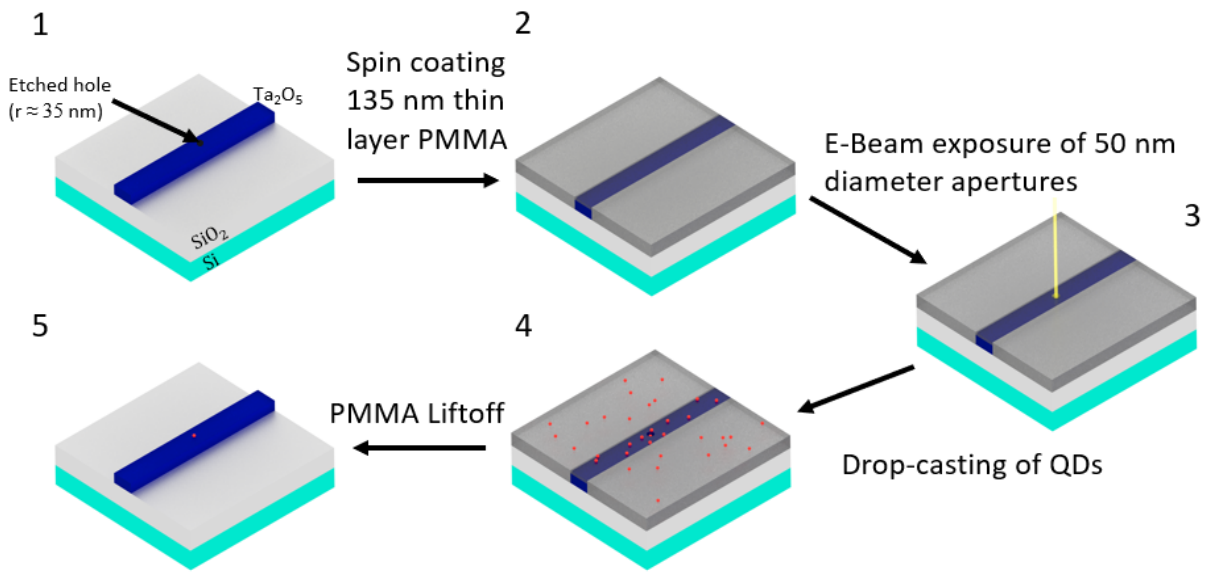

**Figure S3** CQD positioning approach utilizing apertures in polymer thin films. Starting point (1) is a prefabricated chip with $Ta_2O_5$ photonic integrated circuits with a 35 nm radius hole etched into the collection waveguide. Via spin-coating an approx. 135 nm thin PMMA film is applied on the substrate (2). Subsequently, the PMMA film is exposed at desired aperture positions via electron beam lithography (3). Afterwards, the exposed areas are removed during development and a decane solution with diluted CQDs is drop-casted on top of the sample (4). An acetone-based lift-off removes the sacrificial PMMA layer along with the surplus of emitters as a last step (5).

**Experimental system for optical device characterization**

To characterize the fabricated devices, we assess the colloidal quantum dot emitter properties using a home-build confocal microscope as well as fiber-coupled nanophotonic waveguides. In Figure S4 a) a schematic illustration of the confocal microscope configuration is depicted. We use a 532 nm wavelength laser to excite our sample through a microscope objective with a 0.9 numerical aperture. The photoluminescence collected from the sample is separated by a dichroic mirror and recorded with two single-photon avalanche diodes (SPADs) in a Hanbury-Brown-Twiss configuration to assess the statistical properties of the light field. Figure S4 b) illustrates schematically the waveguide-integrated characterization approach. A fiber array is used to couple light into and collect light from the device via multiple channels simultaneously. Channels 1 and 3 are used to collect the photoluminescence and assess the spectral (3) and statistical (1) properties. Channel 2 is used to excite the CQD trough the excitation waveguide with a 532 nm wavelength laser.

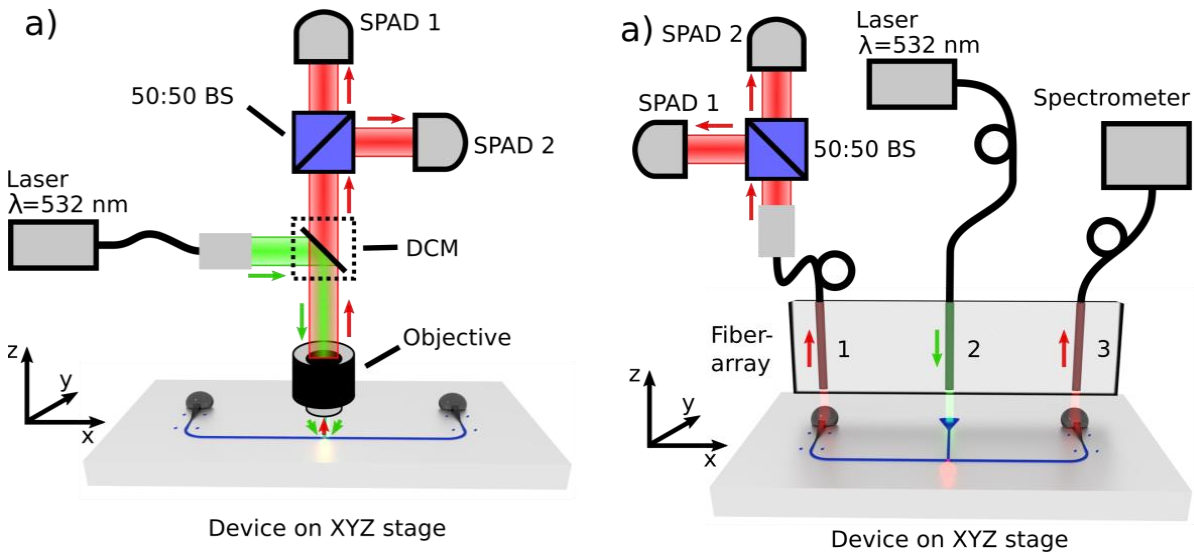

**Figure S4** Optical characterization of waveguide-integrated CQDs. a) Setup for the confocal microscope measurements (BS: beam splitter; SPAD: single-photon avalanche diode; DCM: dichroic mirror). The green arrow indicates the excitation path, the red arrow the collection path. b) Setup for coupling light from / to nanophotonic waveguides on the chip and interface with waveguide-integrated CQDs. The green arrow indicates the excitation path, the red arrow the collection path.

**Measurements of the second-order autocorrelation function**

As a reference, we characterized our devices using the home-build confocal microscope as depicted in Figure S4 a). For a device with suitable design parameters, as that shown in Figure 3 of the main text, we recorded the second-order autocorrelation function for light emitted into the microscope objective using an integration time of 1 hour, which shows antibunching with $g^2(0)_{cm}=0.32 \pm 0.08$ as shown in Figure S5 a). The relatively large variations of the autocorrelation function are a consequence of the low signal levels recorded in confocal microscopy because light from the emitter is predominantly emitted into the waveguide mode, as shown in Fig. S2 a and Fig. 2 f of the main text. Subsequently, we investigated the same device via fiber-coupled nanophotonic waveguides as depicted in Figure 3 b) of the main text. Here, we assess the statistical properties of the waveguide-coupled photoluminescence by again recording the second-order autocorrelation function. Due to the low excitation power and correspondingly low count rate, as well as the blinking behavior, we collected data for approximately 8-12 hours on each device. A fit to the raw data shown in Figure S5 b) yields $g^2(0)_{1nb}=0.43 \pm 0.08$, which reduces to $g^2(0)_{1b}=0.04 \pm 0.08$ when correcting for independently determined background (see Figure 3 c) in the main text). On the same chip we found further devices, which exhibit single CQD occupation. In Figure S5 c) (d)) and e) (f)) we show the recorded second-order autocorrelation functions for two such devices, which show antibunching with $g2(0)_{2b}=0.08 \pm 0.06$ ($g^2(0)_{2nb}=0.55 \pm 0.06$) and $g^2(0)_{3b}=0.08 \pm 0.07$ ($g^2(0)_{3nb}=0.46 \pm 0.07$) after background correction (without background correction), respectively.

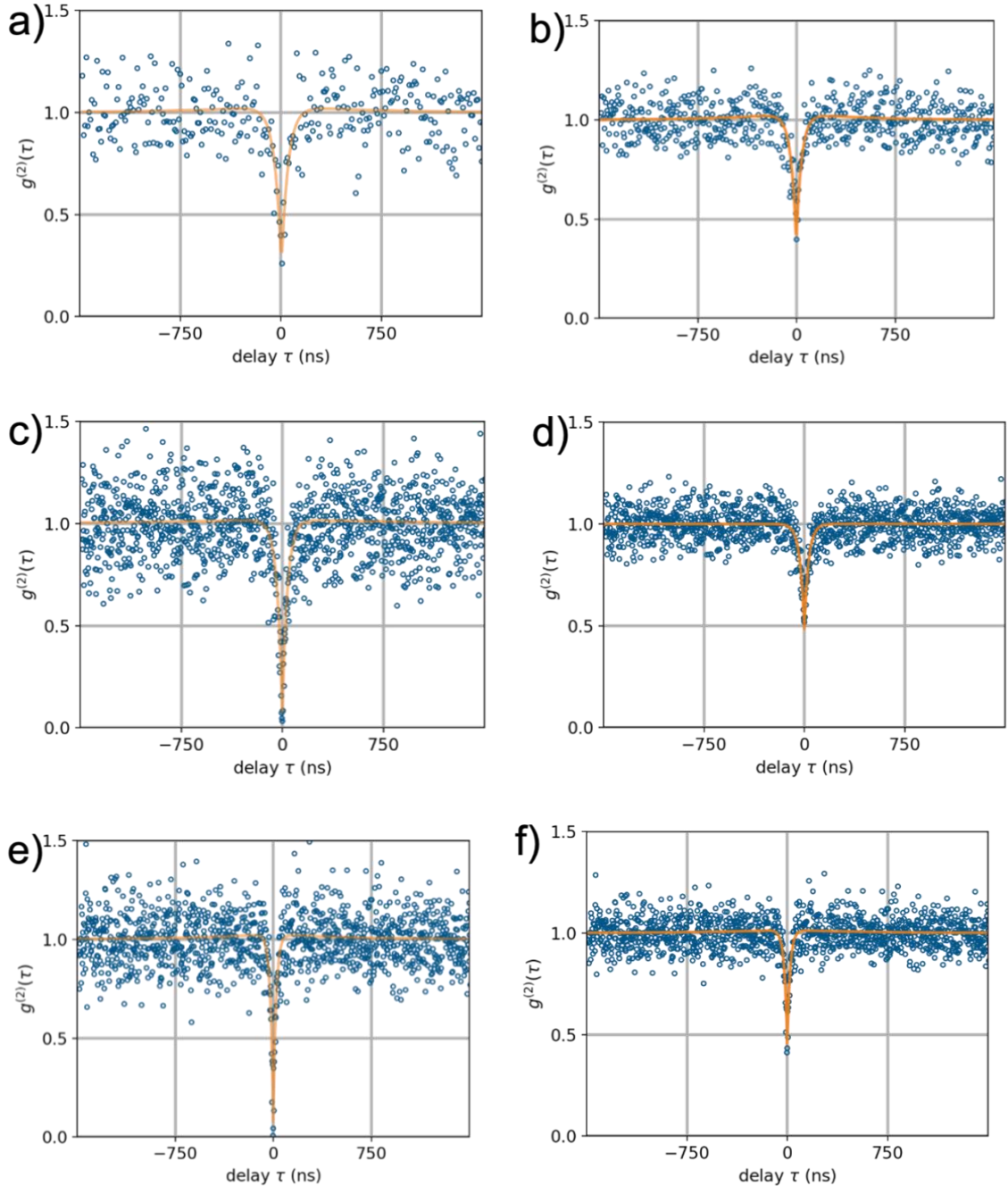

**Figure S5** Second order autocorrelation functions $g^2(\tau)$ for three suitable nanophotonic devices with successfully integrated single CQDs. a) Autocorrelation data recorded in confocal microscopy for the same device as that of Figure 3 in the main text. After correcting for background counts, a fit (orange line) to the data (blue dots) yields $g^2(0)_{cm}=0.32 \pm 0.08$. b) Autocorrelation data for waveguide-coupled photons collected via the optical fiber array for same device as in a) and Figure 3 of the main text. Without correcting for background counts a fit to the data yields $g^2(0)_{1nb}=0.43 \pm 0.08$. Similar devices on the same chip yield c) $g^2(0)_{2b}=0.08 \pm 0.06$ and e) $g^2(0)_{3b}=0.08 \pm 0.07$ after correcting for background noise, corresponding to d) $g^2(0)_{2nb}=0.55 \pm 0.06$ and f) $g^2(0)_{3nb}=0.46 \pm 0.07$ without correcting for background counts.

**Placement yield improvement of colloidal quantum dots on silicon dioxide**

To integrate single CQDs with $Ta_2O_5$ nanophotonic circuits with near unit placement yield, we apply an iterative process: we dropcast CQDs diluted in decane onto a 50 nm thin PMMA film with apertures written at discrete positions in a 5 x 5 pattern on a SiO2-on-Si test chip. Afterwards a lift-off process is performed to remove the PMMA and surplus CQDs. The placement yield of the CQDs for this iteration is determined using the home-build confocal microscope shown in Figure S4 a). We record all occupied sites and find highest yield of 32% for apertures with diameters of 50 nm. For the next iteration (EBL of apertures – dropcast CQDs – lift-off surplus) we eliminate all occupied sites that showed a fluorescence signal from the pattern to be exposed in EBL, such that only empty sites will be exposed. We find that CQDs positioned in earlier iterations are neither harmed by spin-coating PMMA on top of the occupied sites nor by the lift-off process. The iterative process is illustrated in Figure S6. By iterating this process two times, an 80 % placement yield is observed, 48 % of the sites that show a fluorescence signal also showed a second-order autocorrelation function that indicates single photon emission.

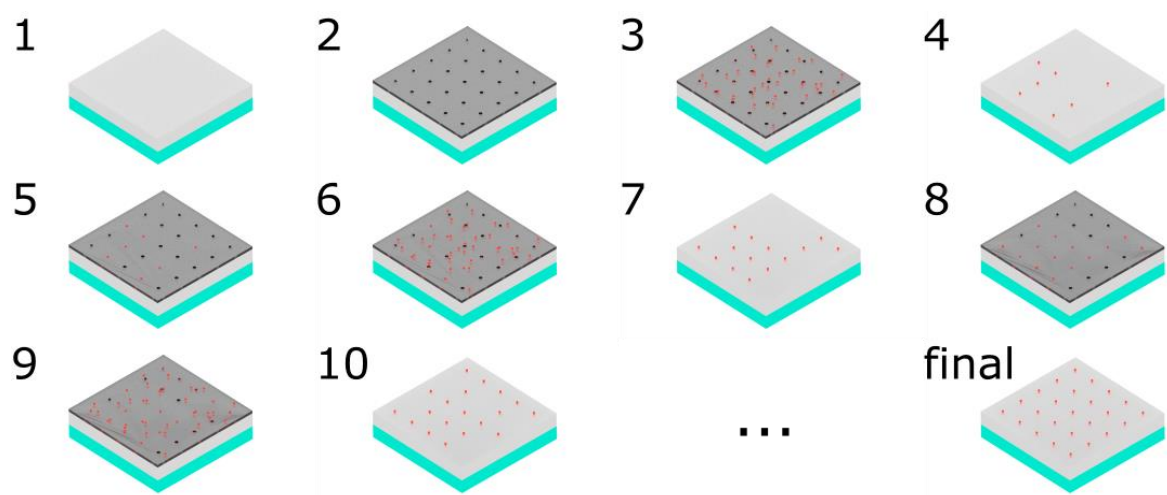

**Figure S6** Iterative CQD placement procedure to improve device yield. Starting point (1.) is a clean SiO2-on-Si sample. A 50 nm thin PMMA film is spin-coated on the substrate and subsequently exposed with a 5x5 pattern consisting of 50 nm holes via EBL. After the development, a decane solution with diluted CQDs is drop-casted on top of the sample. An acetone-based lift-off removes the surplus of emitters along with the sacrificial PMMA layer. The positions, which are filled with CQDs are recorded and steps 2-4 are repeated in 5-7. However, only those sites that did not show appreciable fluorescence in the previous iteration are patterned in this iteration. These steps are repeated (8-10+) until unity placement yield is reached (final).